\pdfoutput=1
\documentclass[%
  aps,
  prb,
  citeautoscript,
  superscriptaddress,
  amsmath,
  amssymb,
  reprint,
  floatfix
]{revtex4-1}

\usepackage[utf8]{inputenc}
\usepackage[T1]{fontenc}
\usepackage{graphicx}
\usepackage{mathtools}
\usepackage[version=4]{mhchem}
\usepackage{textcomp}
\usepackage{braket}
\usepackage[colorlinks=true,urlcolor=blue,linkcolor=blue,citecolor=blue]{hyperref}

\begin{document}

\title{Real-Time Time-Dependent Density Functional Theory Implementation of Electronic Circular Dichroism Applied to Nanoscale Metal-Organic Clusters}

\date{\today}

\author{Esko Makkonen}
\affiliation{Department of Applied Physics, Aalto University, Espoo, Finland}

\author{Tuomas P.\ Rossi}
\affiliation{Department of Physics, Chalmers University of Technology, Gothenburg, Sweden}

\author{Ask Hjorth Larsen}
\affiliation{Simune Atomistics S.L., Donostia/San Sebasti\'an, Spain}

\author{Olga Lopez-Acevedo}
\affiliation{Instituto de F\'{i}sica, Facultad de Ciencias Exactas y Naturales, Universidad de Antioquia, Medell\'{i}n, Colombia}

\author{Patrick Rinke}
\affiliation{Department of Applied Physics, Aalto University, Espoo, Finland}

\author{Mikael Kuisma}
\email{mikael.kuisma@jyu.fi}
\affiliation{Nanoscience Center, University of Jyv\"{a}skyl\"{a}, Jyv\"{a}skyl\"{a}, Finland}

\author{Xi Chen}
\email{xi.6.chen@aalto.fi}
\affiliation{Department of Applied Physics, Aalto University, Espoo, Finland}

\begin{abstract}
Electronic circular dichroism (ECD) is a powerful spectroscopical method for investigating chiral properties at the molecular level. ECD calculations with the commonly used linear-response time-dependent density functional theory (LR-TDDFT) framework can be prohibitively costly for large systems. To alleviate this problem, we present here an ECD implementation for the projector augmented-wave method in the real-time-propagation TDDFT (RT-TDDFT) framework in the open-source GPAW code. Our implementation supports both local atomic basis set and real-space finite-difference representations of wave functions. We benchmark our implementation against an existing LR-TDDFT implementation in GPAW for small chiral molecules. We then demonstrate the efficiency of our local atomic basis set implementation for a large hybrid nanocluster.
\end{abstract}

\maketitle

\section{Introduction}

Chirality is an essential property in several branches of science and technology.
Chiral molecules play a fundamental role in biological activities\cite{Chiral-bio}; for example, DNA double-helices are right-handed and amino acids are left-handed.
Chirality is also critical in pharmaceuticals.
For example, $R$-enantiomer thalidomide is effective against morning sickness for pregnant women, but the $S$-species produce fetal deformations.\cite{Tokunaga,Spek} 
In addition, chiral molecules and chiral nanomaterials have many potential applications in catalysis, sensors, spintronics, optoelectronics, and nanoelectronics \cite{Chiral-cata, Chiral-plasma, Chiral-gold,Chiral-nanotube,chiral-surface, Chiral-gold2,chiral-gold3,chiral-gold4,chiral-gold5,Chiral-asse}.
Therefore, the determination of the handedness of chiral systems is of paramount importance.

Chiral molecules absorb left and right circular polarizations of light differently. This difference is probed in  electronic circular dichroism (ECD) spectroscopy.
ECD is defined as $\Delta \epsilon = \epsilon_L-\epsilon_R$, where $\epsilon_L $ and $\epsilon_R$ are the molar extinction coefficients for left and right circularly polarized light, respectively.
Since ECD is highly sensitive to small details in the atomic structure of molecules and unique for each conformation, it is a powerful technique for characterizing chiral systems and for distinguishing enantiomers\cite{CD-DNA}.

The ECD spectroscopy accompanied by computational modeling can help to get insightful knowledge of the atomic structures of chiral biomolecules and nanoclusters \cite{CD-analysis,XiJCPL,olga}.
For example, in our earlier work, we have identified by comparing simulated and measured ECD that the \ce{Ag+}-mediated guanine duplex has the left-handed helix configuration\cite{XiJCPL}, while the \ce{Ag+}-mediated cytosine has the right-handed helix configuration\cite{olga}.

Most computational ECD approaches are based on the linear-response formalism.\cite{Crawford2006}
Time-dependent density-functional theory (TDDFT) \cite{Runge1984} has become the linear-response method of choice due to its favorable balance of accuracy and computational cost, compared to 
quantum chemical approaches such as coupled cluster and configuration interaction methods.\cite{Crawford2006,Diedrich,Autschbach}
In linear-response TDDFT (LR-TDDFT), the Casida equation\cite{Casida1995,Provorse} is solved in the basis of Kohn--Sham (KS) particle--hole transitions in the frequency domain.\cite{Petersilka,Casida2009}
A full ECD spectrum requires the calculation of a large number of transitions from occupied to unoccupied states.This becomes  computationally prohibitive for large systems with a high density of states, resulting in an unfavorable $O(N^5)$ scaling, where $N$ is the system size. 

An alternative to LR-TDDFT is real-time-propagation time-dependent density-functional theory (RT-TDDFT).
In RT-TDDFT, the system is subjected to an initial perturbation and the KS wave functions are propagated in the time domain  by numerically integrating the time-dependent KS equations.
The real-time approach captures the same information as LR-TDDFT, for small initial perturbations, and incorporates  nonlinear spectral information for larger initial perturbations.\cite{Yabana1,Yabana2, Goings2, Provorse} 

RT-TDDFT scales as $O(N^2)$, but suffers from a large prefactor. LR-TDDFT is usually faster for small systems such as small organic molecules. For large molecules, clusters, and nanoparticles, RT-TDDFT becomes more cost-effective than LR-TDDFT.\cite{Tussipbayev,Provorse}

RT-TDDFT computation of ECD has been implemented for a variety of basis sets:  real-space grids,\cite{Yabana2,Varsano} Gaussian-type atomic orbitals,\cite{Tussipbayev} and a mix of Gaussian-type and plane-wave basis sets\cite{MATTIAT2019110464}. In this work, we present a RT-TDDFT ECD implementation in the open-source GPAW package.\cite{GPAW2, GPAW3} Our implementation uses the projector augmented wave (PAW) method\cite{Blochl1994} and supports both localized basis sets (LCAO mode)\cite{Ask} and real-space grids (grid mode)\cite{GPAW2,GPAW3}. We verify our implementation by comparing our results to those calculated with the existing LR-TDDFT implementation in GPAW. We also benchmark the LCAO mode against accurate real-space grid calculations.

In GPAW, the time-dependent density and potential are expressed on a uniform grid, and the matrix elements of the potential are evaluated on this grid.\cite{GPAW2} The smoothness of these quantities allows for a coarse grid spacing. The LCAO-PAW pseudo wavefunctions can form a local and efficient representation suitable for systems with hundreds of atoms.\cite{Ask} Previous work has shown that LCAO RT-TDDFT in GPAW is capable of simulating the optical spectrum of a silver cluster of more than 500 atoms (\ce{Ag561}).\cite{Kuisma,RosKuiPus17} In this work, we demonstrate the efficiency of our LCAO RT-TDDFT ECD implementation for a large ligand-protected \ce{Ag78} cluster\cite{Ag78} consisting of over 1000 atoms and over 4000 electrons. 

The rest of this paper is organized as follows. In the Methods section we illustrate our implemented methodologies by showing how the ECD is calculated from time-dependent magnetic dipole moment and how the time-dependent magnetic dipole moment is calculated with RT-TDDFT in GPAW. The important information considering all simulations is described in the Computational Methods section. In the Results section we demonstrate the capability of our implementations to predict ECD spectrum for four test cases. Finally, we give a summary in the Conclusion section.

\section{Methods}
\label{Methods}

In RT-TDDFT, the KS wave functions are propagated in time in response to a time-dependent potential starting from an initial state, here chosen to be the ground state. The time-dependent
KS equation is defined as
\begin{equation}
i \frac{\partial}{\partial t}\psi_{n}({\bf r},t)=H_\text{KS}(t)\Psi_{n}({\bf r},t),
    \label{KSTDDF}
\end{equation}
where $H_\text{KS}(t)$ is the time-dependent KS Hamiltonian and $\psi_{n}({\bf r},t)$ is a time-dependent KS wave function.
A common practice in time-propagation schemes is to use the weak $\delta$-kick approach \cite{Yabana1} to calculate the linear-response functions.
After perturbing $H_\text{KS}(t)$ by the $\delta$-kick at $t = 0$,  Eq.~\eqref{KSTDDF} is propagated using the semi-implicit Crank-Nicolson method, whose numerical reliability  has been demonstrated previously.\cite{Kuisma}

In this work, we implement and thoroughly benchmark the calculation of time-dependent magnetic moment within the time-propagation framework for obtaining the ECD spectrum.
In the following, we derive the relevant equations within the PAW method \cite{Blochl1994}.
This derivation partly follows the one shown by Varsano \textit{et al.} \cite{Varsano}

\subsection{ECD from the Induced Time-Dependent Magnetic Moment}

A commonly used experimental quantity to measure ECD is the difference in molar extinction coefficients, which is given by
\begin{align}
    \Delta \epsilon(\omega) &= \frac{16\pi N_\text{A}}{3\log(10)10^3}\frac{2\pi}{\hbar c} \omega R(\omega)^\text{cgs}.
    \label{epsilon}
\end{align}
Here $\omega$ is the the energy of the incident light, $c$ the speed of light, $\hbar$ the reduced Planck constant, $N_\text{A}$  Avogadro's constant, and $R(\omega)^\text{cgs}$ the rotatory strength in cgs units. The quantity that characterizes $\Delta\epsilon(\omega)$ and therefore the ECD spectrum is the rotatory strength. The relationship between rotatory strength in cgs units and rotatory strength in atomic units (denoted as $R(\omega)$) is
\begin{align}
    R(\omega)^\text{cgs}&= \frac{e^2\hbar^2}{m_e\alpha} 10^6 R(\omega),
    \label{RotUnits}
\end{align}
where $e$ is the elementary charge, $m_e$ the mass of an electron, and $\alpha$ the fine structure constant. We will work in
atomic units and perform the required unit conversions afterwards.

The rotatory strength is defined through the optical rotatory response tensor by
\begin{equation}
    R (\omega)=\frac{\omega}{\pi c} \mathrm{Im}\bigg[\sum_{k} \beta_{kk}(\omega)\bigg],
    \label{Rot}
\end{equation}
where index $k$ enumerates Cartesian  coordinates ($k\in \{x,y,z\}$). 
Next we will derive the relationship between the optical rotatory response tensor and the time-dependent magnetic dipole moment, which is the quantity calculated in TDDFT.

For a system in an external electric field $\mathbf{E} = [E_x, E_y, E_z]$ (no external magnetic field present), the induced time-dependent magnetic dipole moment in direction $j$, $m_{j}(t)$, has the expansion
\begin{align}
    m_{j}(t) &=
    \frac{1}{c}\sum_k \int_{-\infty}^{\infty}\beta_{jk}(t-\tau)\frac{\partial E_{k}(\tau)}{\partial \tau} {\rm d}\tau\nonumber\\ 
    &+ \text{higher-order terms}
     \label{eq2}
\end{align}
where $\beta_{jk}$ is the $jk$ component (indices $j,k\in \{x,y,z\}$) of the optical rotatory response tensor, and $E_{k}$ is the electric field component. The response tensor $\beta_{jk}$ describes the induced magnetic dipole moment in the $j$ direction for a perturbing electric field in the $k$ direction. The response is causal, which means that $\beta_{jk}(t)$ vanishes for negative values of $t$.

In the weak-field limit, the time-dependent magnetic moment is dominated by the first-order term given by the linear-response functions $\beta_{jk}(t)$.
By using the convolution theorem and the properties of Fourier transforms, the first-order term of Eq.~\eqref{eq2} can be written in the frequency domain as
\begin{align}
    m_{j}(\omega) &= -\frac{i\omega}{c}\sum_k \beta_{jk}(\omega)E_{k}(\omega).
    \label{eq4}
\end{align}
To resolve all components of $\beta_{jk}(t)$, we perform the $\delta$-kick\cite{Yabana1} in all three Cartesian directions using a perturbing electric field of the form $\mathbf{E}^{(k)}(t) = \kappa \hat {\bf k} \delta(t)$. The  $k$ superscript in parenthesis indicates the kick direction, to be distinguished from the component subscript. 
We keep the intensity $\kappa$ weak to restrict our calculations to the linear-response regime.

In the frequency domain, the delta kick becomes a constant for all frequencies
\begin{equation}
E^{(k)}_{k'}(\omega) = \kappa \delta_{kk'}.
\label{efield}
\end{equation}
Eq.~\eqref{eq4} then simplifies to
\begin{align}
    \beta_{jk}(\omega) = \frac{ic}{\kappa \omega} m^{(k)}_{j}(\omega).
    \label{bjk}
\end{align}
Combining Eqs.~\eqref{Rot} and \eqref{bjk}, we get the equation for rotatory strength expressed  with magnetic dipole moment
\begin{equation}
    R (\omega)=\frac{1}{\pi\kappa} \mathrm{Re}\bigg[\sum_{k} m_{k}^{(k)}(\omega)\bigg].
    \label{RotwithmREAL}
\end{equation}
In our methodology, $m^{(k)}_{j}(\omega)$ is calculated by Fourier transform of $m^{(k)}_{j}(t)$, which has been obtained through time-propagation,
\begin{equation}
    m^{(k)}_{j}(\omega) = \int_{0}^{\infty} e^{i\omega t} m^{(k)}_{j}(t)~{\rm d}t.
    \label{asd}
\end{equation}
In principle, the integration interval goes from zero to infinity.
In practice, a finite propagation time ($T$) suffices by introducing an artificial lifetime $\omega \rightarrow \omega + i\frac{\sigma^2}{2}t$, where $\sigma$ is the parameter that determines the line width of the Gaussian line shape. Introducing this into Eq.~\eqref{asd} gives   
\begin{equation}
    m^{(k)}_{j}(\omega) = \int_{0}^{T} e^{i\omega t} e^{-\frac{\sigma^2}{2}t^2} m^{(k)}_{j}(t)~{\rm d}t.
    \label{m_gauss}
\end{equation}
For a desired value of $\sigma$, the propagation time $T$ needs to be large enough so that  $e^{-\frac{\sigma^2}{2}T^2} \approx 0$.

\subsection{Computing the Magnetic Moment \texorpdfstring{$m^{(k)}_{j}(t)$}{} in the PAW formalism}

The magnetic moment is defined by the following operator (in atomic units)
\begin{align}
  \hat{\mathbf{m}} = -\frac{i}{2c} \hat{\mathbf{r}} \times \hat{\nabla}.
\end{align}
The expectation value of the time-dependent magnetic moment is obtained as
\begin{align}
  \mathbf{m}(t) = \sum_{n} f_n \int \psi_n^*({\bf r},t) \,\hat{\mathbf{m}}\, \psi_n({\bf r},t)~{\rm d}\mathbf{r},
  \label{expecgridm2}
\end{align}
where $f_n$ is the occupation number of the $n$:th KS state and $\psi_n({\bf r},t)$ is the time-evolved KS wave function.

In the PAW method \cite{Blochl1994} the wave functions $\psi_n({\bf r},t)$ have decomposition
\begin{equation}
   \psi_n({\bf r},t) = \tilde\psi_n({\bf r},t) +\sum_{ai} \Big[\phi_i^a ({\bf r})-\tilde\phi_i^a ({\bf r})\Big]\braket{\tilde p_i^a|\tilde\psi_n(t)},
 \label{PAW}
\end{equation}
where
\emph{$\tilde\psi_n({\bf r},t)$} is a smooth \emph{pseudo wave function} and \emph{$\sum_{ai} \Big[\phi_i^a ({\bf r})-\tilde\phi_i^a ({\bf r})\Big]\braket{\tilde p_i^a|\tilde\psi_n(t)}{}$} a local correction inside an atomic \emph{augmentation sphere}. \emph{$\tilde p_i^a$} is a localized \emph{projector function} and \emph{$\phi_i^a$} and \emph{$\tilde\phi_i^a$} are \emph{partial} and \emph{pseudo partial waves}, respectively. These quantities are specific to PAW.
In the PAW formalism, the expectation value in Eq.~\eqref{expecgridm2} becomes
\begin{align}
  \mathbf{m}(t)=&\sum_n f_n \braket{\tilde\psi_n(t)|\hat{\mathbf{m}}|\tilde\psi_n(t)}{}\nonumber\\
 & +\sum_{naij}f_n\braket{\tilde\psi_n(t)|\tilde p_i^a}{}\Delta {\bf M}_{ij}^a\braket{\tilde p_j^a|\tilde\psi_n(t)}{},\label{withM}
\end{align}
where the augmentation-sphere contribution is  $\Delta {\bf M}_{ij}^a = \braket{\phi_i^a|\hat{\mathbf{m}}|\phi_j^a} - \braket{\tilde\phi_i^a|\hat{\mathbf{m}}|\tilde\phi_j^a}$.
For evaluating $\Delta {\bf M}_{ij}^a$, the required matrix elements of the form
$\braket{\phi_i^a|\mathbf{r} \times \nabla|\phi_j^a}$
are evaluated in two atom-centered parts as
$\braket{\phi_i^a|(\mathbf{r} - \mathbf{R}^a) \times \nabla|\phi_j^a} + \mathbf{R}^a \times \braket{\phi_i^a|\nabla|\phi_j^a}$,
where $\mathbf{R}^a$ is the coordinate of atom $a$.

In the LCAO expansion, the time-dependent pseudo wave function $\tilde\psi_n({\bf r},t)$ is written as a linear combination of atom-centered basis functions $\varphi_{\mu}(\mathbf{r}-\mathbf{R}^a)$
\begin{equation}
\tilde\psi_n({\bf r},t) = 
\sum_\mu c_{\mu n}(t) \varphi_{\mu}(\mathbf{r}-\mathbf{R}^a),
\label{atombasis}
\end{equation}
where $c_{\mu n}(t)$ are the time-dependent expansion coefficients.
With this LCAO expansion, Eq.~\eqref{withM} can be written compactly as
\begin{align}
  \mathbf{m}(t)
  &= \sum_{\mu\nu} \rho_{\nu\mu}(t){\bf M}_{\mu\nu},
  \label{mLCAO}
\end{align}
where $\rho_{\mu\nu}(t)=\sum_n f_n c_{\mu n}(t)c_{\nu n}^*(t)$ is the KS density matrix in the LCAO basis. The matrix elements $\bf M_{\mu\nu}$ are given by the pseudo and augmentation contributions 

\begin{align}
  \bf M_{\mu\nu} &= \tilde{\mathbf{M}}_{\mu\nu} + \Delta\bf  M_{\mu\nu} \\
  &\nonumber\\
  \tilde{\mathbf{M}}_{\mu\nu} &= \braket{\varphi_\mu|\hat{\mathbf{m}}|\varphi_\nu} \\
  &\nonumber\\
  \Delta\bf M_{\mu\nu} &= \sum_{aij} \braket{\varphi_\mu|\tilde p_i^a} \Delta {\bf M}_{ij}^a \braket{\tilde p_j^a|\varphi_\nu}.
\end{align}

In real-space grid mode the magnetic moment is calculated using equation Eq.~\eqref{withM}. In LCAO mode, Eq.~\eqref{mLCAO} is used. The matrix elements $\bf M_{\mu\nu}$ are time independent and calculated only once before the time propagation.

After a complete time propagation, the recorded $\mathbf{m}(t)$ is transformed to frequency domain as a post processing step according to Eq.~\eqref{m_gauss} at each desired $\omega$ value. Then the rotatory strength is calculated according to Eq.~\eqref{RotwithmREAL}.

We note that research of gauge origin issues is not within the scope of this work and the reader is suggested to explore a recent detailed investigation on the matter.\cite{MATTIAT2019110464}

\subsection{Computational Methods}
\label{Computational Methods}

For our calculations in this work, we used the PBE exchange-correlation functional\cite{PBE}, unless otherwise mentioned. The molecules were placed into a cubic unit cell with the vacuum size of 8~\AA. The real-space grid spacing was chosen as $h=0.2$~\AA. We tested coarser settings with $h=0.3$~\AA~in RT/LCAO mode for the \ce{Ag+}-mediated guanine duplex case to demonstrate that such a coarser grid is sufficient for calculating the ECD spectrum within LCAO mode, where the grid is used to represent only real-space density and potential.\cite{Ask,Kuisma}

Per atom, the electronic configuration of valence electrons is H(1s$^1$) O(2s$^2$2p$^4$), C(2s$^2$2p$^2$), N(2s$^2$2p$^3$), S(3s$^2$3p$^4$), P(3s$^2$3p$^3$), F(2s$^2$2p$^5$) and Ag(4d$^{10}$5s$^1$). The remaining electrons were treated as a frozen core. 
The default PAW dataset package 0.9.20000 was used for all the atoms.

In the LCAO mode, the default GPAW double-zeta polarized (dzp) basis sets\cite{Ask} were used for all other elements, unless otherwise mentioned. For Ag, the optimized double-zeta basis set (so-called "p-valence" basis set) was used for Ag atoms.
In this basis set, the default p-type polarization function is replaced with a bound unoccupied p-type orbital and its split-valence complement. The inclusion of 5p orbitals in the valence improves the chemistry and photochemistry as showed in a previous work.\cite{Kuisma}

To test the effects of basis set in ECD simulations, more complete basis sets were constructed by adding diffuse augmentation functions
through truncated numerical Gaussian-type orbitals (NGTOs) to the default dzp basis sets.\cite{Rossi2015}
We denote these basis sets  dzp+NGTOs.
Our approach follows a recent study of introducing augmentation functions that demonstrated good results for Bethe-Salpeter equation (BSE) and LR-TDDFT calculations for  molecules with numeric atom-centered orbitals.\cite{aims-aug} 
Gaussian basis function exponential parameters, the $\zeta$-parameters,  were taken from aug-cc-pvdz basis sets tabulated in Basis Set Exchange.\cite{BSE1,BSE2} The parameters are tabulated in Supplementary Table~S1.

For comparison, we also calculated the ECD by existing LR-TDDFT methods in GPAW. The LR-TDDFT approach of GPAW chooses a cut-off for the Kohn-Sham single-particle excitations and diagonalizes the Casida matrix, hence there exists a cut-off parameter in these calculations.  We choose high cut-off ($>$ 20 eV)  to compare with our RT-TDDFT results in this work. The effect of the cut-off to the convergence of LR-TDDFT is discussed in Supplementary Note~S1.

An artificial life-time  for  the electron dynamics was introduced via Gaussian line shape with $\sigma=0.2$ eV in all figures unless otherwise mentioned. In this work, the rotatory strength is presented in units $10^{-40}\ \text{erg}\cdot \text{esu}\cdot \text{cm}\cdot \text{Gauss}^{-1} \text{eV}^{-1}=10^{-40}\ \text{cgs}\ \text{eV}^{-1}$.

The reported computational run times are obtained with Intel Xeon Gold 6230 processors with Mellanox HDR InfiniBand interconnect as installed in Puhti supercomputer at CSC -- Finnish IT Center for Science. To support open data-driven chemistry and materials science \cite{Himanen/Geurts/Foster/Rinke:2019}, we will upload all calculations of this work to the Novel Materials Discovery (NOMAD) laboratory and open-access Zenodo repository.

\section{Results}
\label{Results}

In this section, we present four test cases for our implementation. First, we use a benchmark molecule ($(R)$-methyloxirane) to validate that our RT-TDDFT implementation can produce the same ECD spectra as the LR-TDDFT implementation in both LCAO and real-space grid mode. Then, we use a chiral \ce{Ag4} string and a \ce{Ag+}-mediated guanine duplex G$_2-$Ag$_2^{2+}-$G$_2$ structure \cite{XiJCPL,IJMS} to demonstrate that LCAO RT-TDDFT adequately reproduces the rotatory strength of the reference grid mode calculation up to 8 eV. Finally, we apply the LCAO RT-TDDFT approach to a  hybrid silver cluster
$[$\ce{Ag78}($S$-BDPP)$_6$(SR)$_{42}]$ (hereafter denoted as \ce{Ag78}), where BDPP = 2,4-bis-(diphenylphosphino)pentane and SR = SPhCF$_3$ \cite{Ag78}. The whole system is considerably large for TDDFT simulations.
We will show that LCAO RT-TDDFT is computationally efficient and produces  ECD spectra that compare well with  experimental results.

\begin{figure}[b!]
  \centering
  \includegraphics[width=0.7\linewidth]{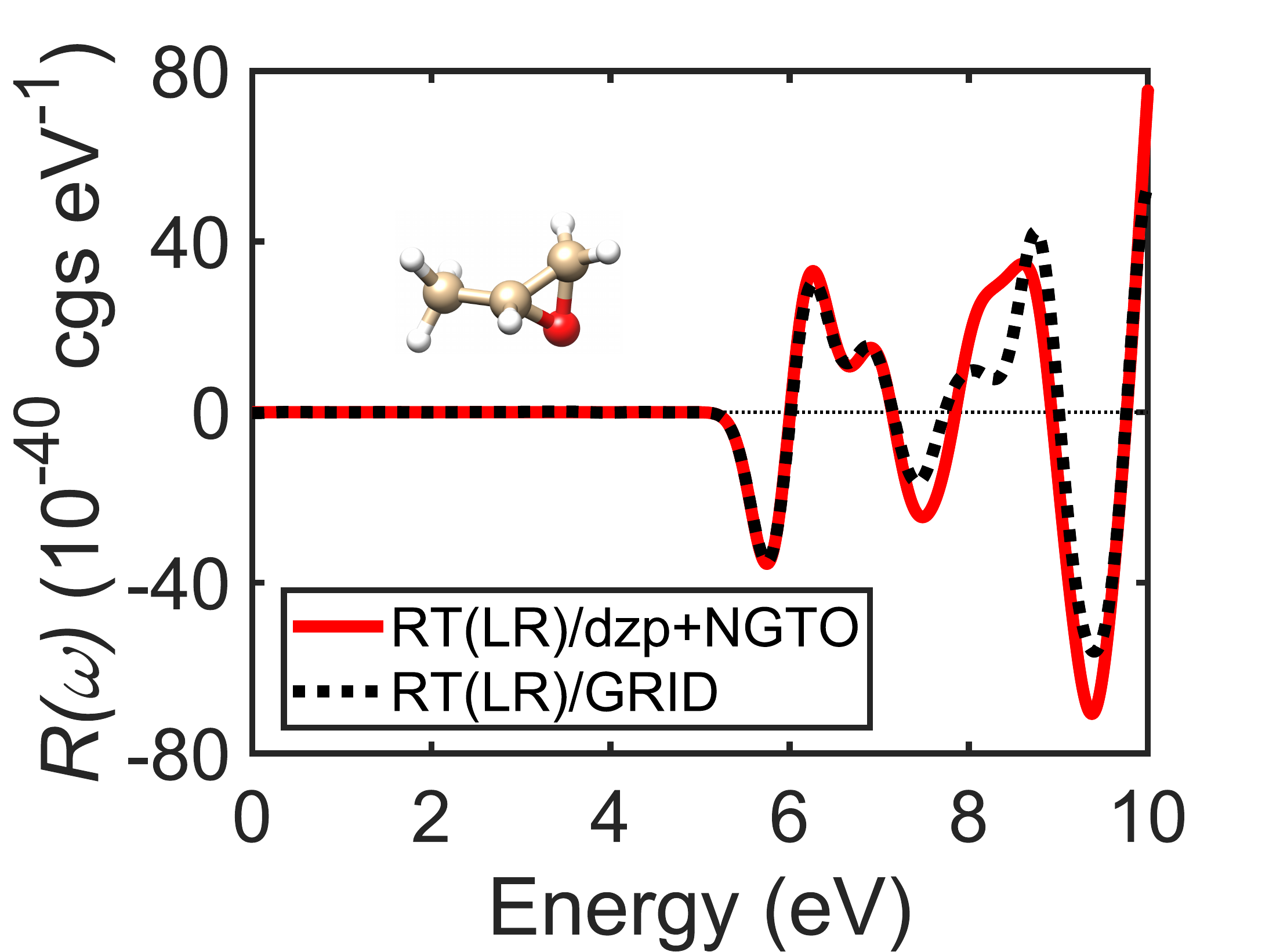}
  \caption{Rotatory strength of $(R)$-methyloxirane calculated by RT-TDDFT and LR-TDDFT in both LCAO and real-space grid modes.}
  \label{fig:rmeth2inone}
\end{figure}

\begin{figure*}[t!]
  \centering
  \includegraphics[width=1\linewidth]{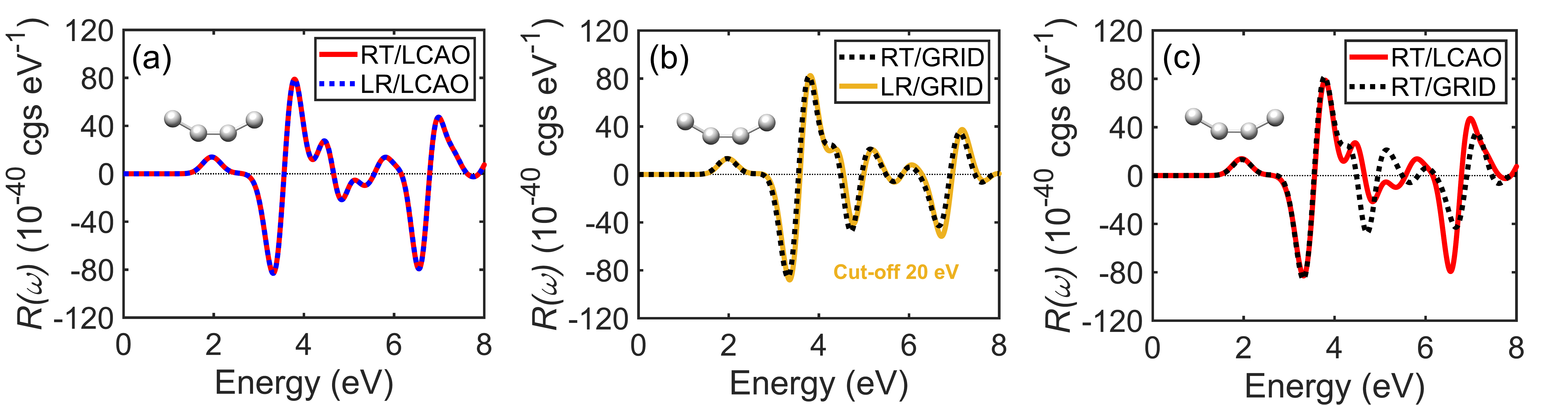}
  \caption{Rotatory strength of a \ce{Ag4}-string (inset figures) calculated with (a)~the LCAO mode and (b)~the grid mode. (c)~Comparison between the two modes.}
  \label{fig:Ag4string4-2}
\end{figure*}

\subsection{\texorpdfstring{$(R)$}{(R)}-methyloxirane}
\label{r-methyloxirane}

$(R)$-methyloxirane is one of the most typical benchmarks for optical activity calculations.\cite{Varsano,MATTIAT2019110464} Therefore, we choose this chiral molecule as our first test system. The atomic structure is taken from the NIST database. The structure was optimized with the configuration interaction singles-doubles (CISD) method and a 6-31G* Gaussian orbital basis set. \cite{str}

The ECD of $(R)$-methyloxirane was calculated both with our RT-TDDFT implementation and LR-TDDFT. All RT-TDDFT calculations were propagated to  $T=30$~fs in steps of 5~as.  The dzp+NGTO basis set was used in the LCAO simulations. The RT-TDDFT spectra look identical to the LR-TDDFT ones in both LCAO and real-space grid mode (the ECD is shown separately for LCAO and real-space grid cases in Supplementary Figure~S1). The maximum difference is less than 0.5 in cgs units. The dzp+NGTO accurately predicts the four first peaks in comparison to real-space grid calculation, as shown in Figure \ref{fig:rmeth2inone}, but the dzp basis doesn't give accurate results in this case (Supplementary Figure~S2).

\subsection{\texorpdfstring{\ce{Ag4}}{Ag4} string}
\label{ag4-string}

To test our method on metallic systems, we use a chiral silver string as the second example. 
The \ce{Ag4} string was built artificially, with a bond length of 2.7~{\AA}, Ag-Ag-Ag angle of 150$^{\circ}$, and torsion angle of 10$^{\circ}$ as shown in the inset of Figure~\ref{fig:Ag4string4-2}.
In the RT-TDDFT simulations, the propagation was carried out up to $T=30$~fs in steps of 5~as.

Figure \ref{fig:Ag4string4-2}a and Figure \ref{fig:Ag4string4-2}b show that our RT-TDDFT implementation again successfully reproduces the rotatory strength of  LR-TDDFT both in LCAO and grid mode.
The slight disagreement of LR-TDDFT and RT-TDDFT in grid mode above 4~eV is due to a  difficult LR-TDDFT convergence, which is discussed more in detail in Supplementary Note~S1.

Figure \ref{fig:Ag4string4-2}c shows that LCAO again adequately reproduces the rotatory strength from the more accurate grid mode up to 8~eV, which is
higher than energies commonly used for recording experimental spectra.

\subsection{\texorpdfstring{Ag${^+}$}{Ag+}-mediated guanine duplex}
\label{ag-guanine}

\begin{figure}[t!]
  \centering
  \includegraphics[width=0.8\linewidth]{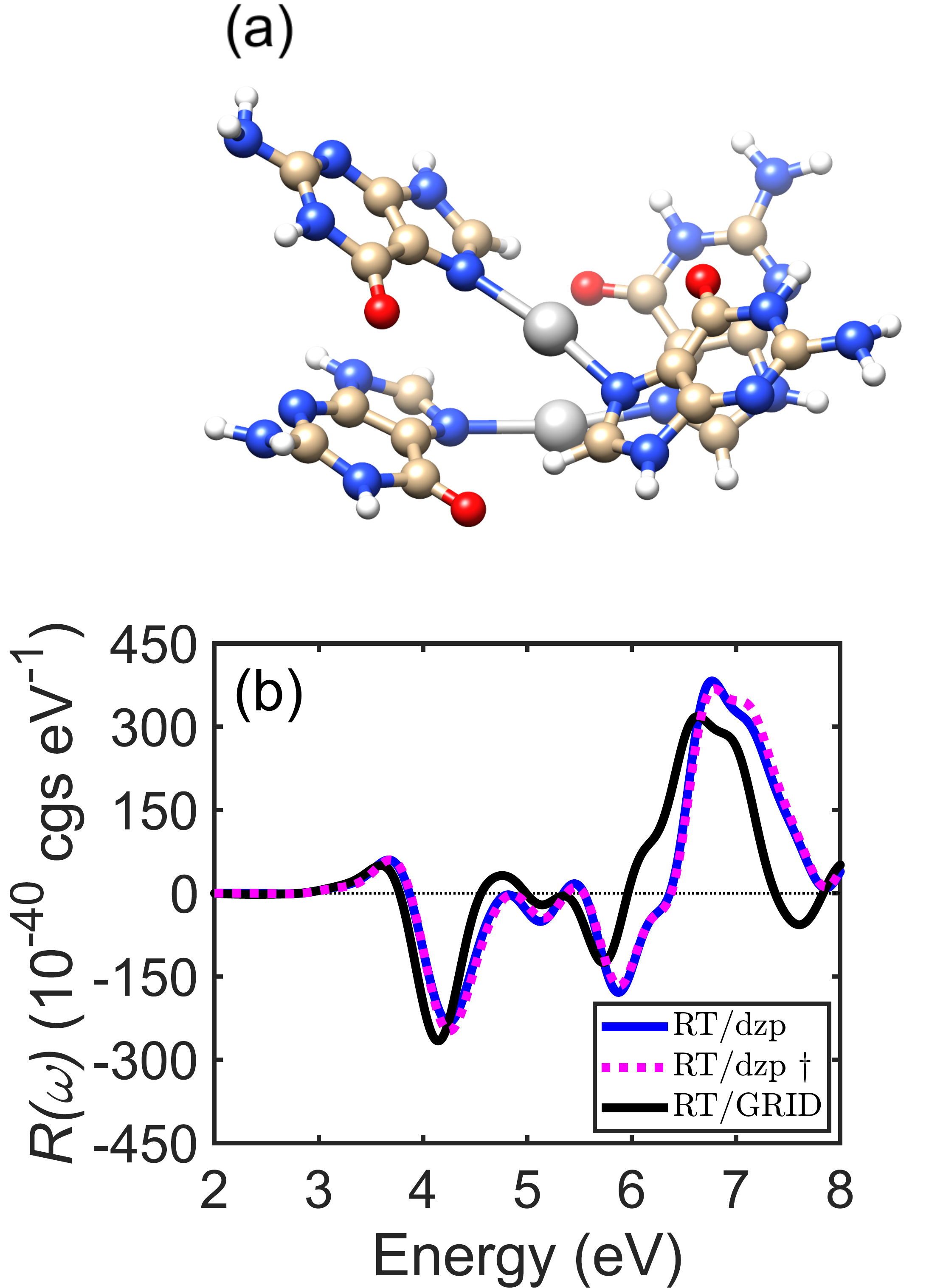}
 \caption{(a)~Structure and (b)~rotatory strength of  G$_2-$Ag$_2^{2+}-$G$_2$.
 LCAO calculations use dzp basis sets.
 Here $\dagger$ notes settings with grid parameter $h=0.3$~\AA, and propagation of 30~fs in steps of 10~as.}
  \label{fig:GG2nosug2}
\end{figure}

After testing on a molecule and a silver string, we apply our method to an organic-metal hybrid system. 
We use one configuration of the \ce{Ag+}-mediated guanine duplex (G$_2-$Ag$_2^{2+}-$G$_2$, Figure~\ref{fig:GG2nosug2}a) from our previous work \cite{XiJCPL}. The purpose here is to benchmark the accuracy of the LCAO method in a more complex system.

Comparing the results calculated with the two modes in Figure~\ref{fig:GG2nosug2}b, we find that the dzp basis set reproduces almost the same rotatory strength up to 6~eV, covering the energy window of most experimentally measured ECD spectra. The dzp+NGTO basis improves the agreement up to 8 eV as shown in Supplementary Figure~S3.

The benefit of the LCAO mode is its low computational cost. For this system, the LCAO mode is over 10 times faster (9 hours on 80 cores versus 40 hours on 240 cores).
Furthermore, we calculated the ECD with a coarser grid $h=0.3$~Å to represent real-space density and potential and a larger time step 10~as.
The ECD lies on top of the one obtained from previous RT/dzp as shown in Figure~\ref{fig:GG2nosug2}b.
The simulation with coarser parameters took only 2.5 hours using 80 cores, which is about 50 times faster than the grid mode.

\subsection{\texorpdfstring{\ce{Ag78}}{Ag78} cluster}
\label{ag78cluster}

\begin{figure}[b!]
  \centering
  \includegraphics[width=1\linewidth]{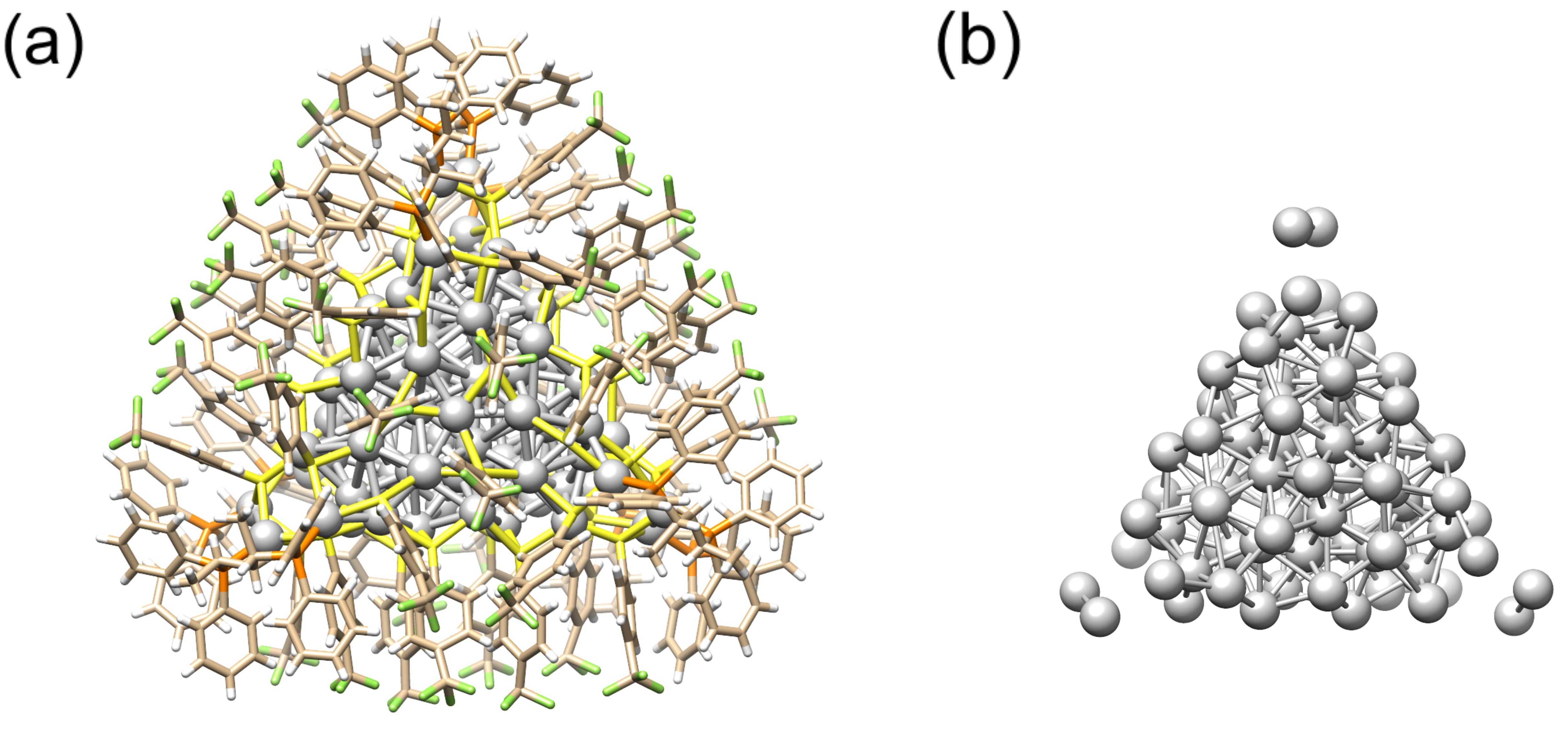}
  \caption{(a) The structure of the \ce{Ag78} cluster. Ligand atoms: H, white; C, beige; F, green; P, orange; S, yellow. (b) Ag atoms in the \ce{Ag78} cluster.}
  \label{fig:Ag78str} 
\end{figure}

In this test case, we illustrate the efficiency and accuracy of our RT-TDDFT/LCAO methodology on a ligand-protected \ce{Ag78} cluster (Figure~\ref{fig:Ag78str}a)  and present a comparison between the experimentally measured\cite{Ag78} and the calculated ECD spectra.
We used the X-ray structure \cite{Ag78} in calculations.
The 78 silver atoms (Figure \ref{fig:Ag78str}b) are covered by ligand molecules containing C, N, O, F, H and S atoms (Figure \ref{fig:Ag78str}a).
The total number of atoms is 1074 and the number of valence electrons is 4272.
The large size and the complexity of the cluster make it an ideal system to test the computational efficiency of the RT-TDDFT/LCAO approach. 

\begin{figure}[t!]
  \centering
  \includegraphics[width=1\linewidth]{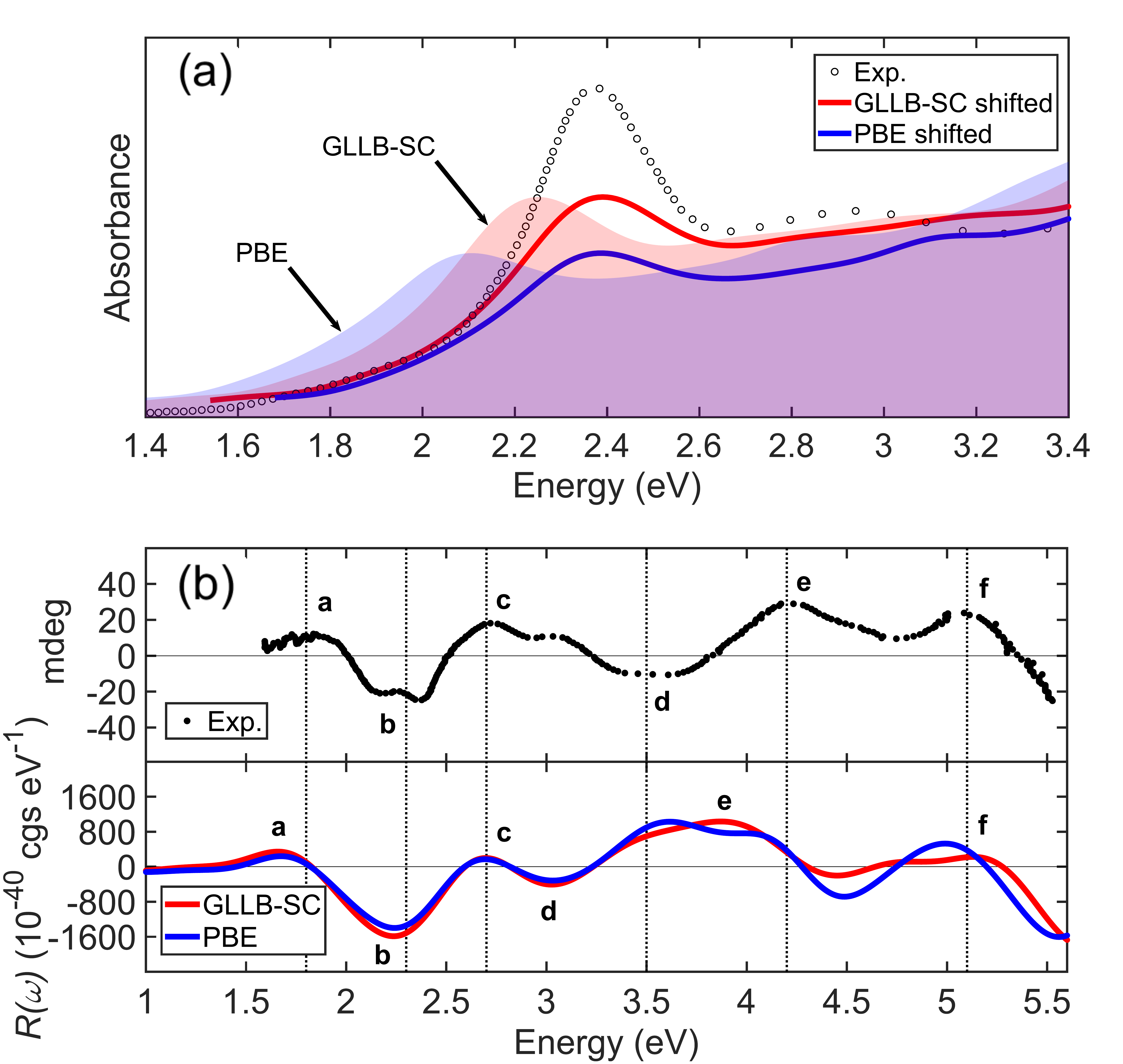}
  \caption{(a) Photoabsorption spectrum of \ce{Ag78}. Shaded areas represent calculated spectra and lines shifted calculated spectra. The GLLB-SC spectrum was shifted by 0.14 eV and the PBE spectrum by 0.28 eV. Gaussian broadening with $\sigma=0.1$ eV was applied. (b) The experimental (top panel) and calculated (lower panel) ECD spectra of \ce{Ag78}. Default dzp basis sets were used for other than Ag atoms. Gaussian broadening with $\sigma=0.2$ eV was applied to approximately match the
spectral linewidth with the experimental data. The calculated ECD spectra were shifted according to the shifts done for calculated absorption spectra.}
  \label{fig:Ag78EXPCALC} 
\end{figure}

Due to its $O(N^5)$ scaling LCAO LR-TDDFT was not applicable to the \ce{Ag78} cluster with our computational resources. However, the scaling of LCAO RT-TDDFT is only $O(N^3)$, which made it possible to calculate ECD spectra for \ce{Ag78}.

In addition to the PBE exchange-correlation functional, we also used the GLLB-SC exchange-correlation potential\cite{KuiOjaEnk10}. 
The GLLB-SC functional was chosen because  former  studies  show that  the GLLB-SC functional provides more accurate predictions of  the  optical  absorption  spectra  of  Ag  clusters  with respect to both the local density approximation (LDA) and the  generalized  gradient  approximations  (GGA)\cite{Kuisma}. The real-time propagation was taken up to $T= 30$~fs  in 10~as steps and a grid spacing of $h= 0.3$~Å was used.

For the \ce{Ag78} cluster,  we have also calculated the photoabsorption spectrum with LCAO RT-TDDFT (Figure~\ref{fig:Ag78EXPCALC}a). Both GLLB-SC and PBE reproduce the first peak of the measured absorption spectrum.  However, the GLLB-SC spectrum is red-shifted by 0.14~eV and PBE by 0.28~eV. 

Table \ref{tbl:table1} and Figure \ref{fig:Ag78EXPCALC}b  present the comparison between the experimentally measured\cite{Ag78} and the calculated ECD spectra. The TDDFT spectra are shifted to higher energies by the same amount as the optical spectra in  Figure \ref{fig:Ag78EXPCALC}a).
Both the PBE and GLLB-SC functional capture the main features of the experimental ECD, which are the four positive peaks (a, c, e, f) and the two negative peaks (b, d).

We now briefly discuss the differences between the theoretical spectra and the experimental spectrum. The calculated absorption and ECD spectra are shifted to lower energies most likely due to the underestimation of the energy gap between occupied and unoccupied KS states in the DFT simulations and mismatches in the Ag d-band location.
The underestimation is less pronounced in the GLLB-SC calculations, because GLLB-SC introduces an orbital-energy dependent localization of the exchange hole and describes Ag d-orbitals more accurately. However, the improved description of the energy gap in GLLB-SC does not remove the mismatch of peaks d and e, suggesting there may be transitions that need a better description. 
Furthermore, the ECD spectrum was measured in a solvent. The fact that our calculations are performed for the experimental crystal X-ray structure and without  conformational sampling may contribute to the differences.
Using the dzp+NGTO basis set does not remove these differences as demonstrated in Supplementary Figure~S4.

The calculation of \ce{Ag78} system took 24 hours with 200 cores with the PBE exchange-correlation functional and 33 hours with the GLLB-SC functional.
This is remarkably fast for TDDFT ECD calculations of such a large system.

\begin{table}
  \caption{The ECD peak positions in Figure \ref{fig:Ag78EXPCALC} (in unit eV).}
  \label{tbl:table1}
  \begin{tabular}{cccc}
    \hline
    \textbf{Peak}  & \textbf{Experiment} & \textbf{PBE} &\textbf{GLLB-SC} \\
    \hline
    \textbf{a} & 1.8 & 1.68 & 1.67\\
    \textbf{b} & 2.3 & 2.23 & 2.27\\
    \textbf{c} & 2.7 & 2.69 & 2.69\\
    \textbf{d} & 3.5 & 3.04 & 3.03\\
    \textbf{e} & 4.2 & 3.83 & 3.79\\
    \textbf{f} & 5.1 & 4.99 & 5.14\\
    \hline
  \end{tabular}
\end{table}

\section{Conclusions}
\label{conclusion}

We present a RT-TDDFT implementation for calculating ECD in GPAW package, which supports both LCAO mode and grid mode. 
While RT-TDDFT/LCAO is less accurate than RT-TDDFT/GRID, our tests have shown that the LCAO method nevertheless produces matching spectra in the experimentally relevant energy ranges. 

The high computational efficiency of the RT-TDDFT/LCAO is enabled by the combination of localized orbitals and the PAW method.
We demonstrated the efficiency of our code by computing ECD spectra of a large hybrid nanocluster with thousand atoms, a system whose ECD is challenging to compute by RT-TDDFT/GRID or conventional linear-response formalisms.

Our RT-TDDFT implementation with localized orbitals and PAW in GPAW opens the door to study the large-scale chiral systems with good accuracy and efficiency.
We expect that our open-source implementation will be advantageous for studying the chiroptical property of large systems without excessive computational cost, which will help to develop many chirality related applications.

\begin{acknowledgments}
This work was supported by the Academy of Finland, Projects 308647, 314298, 279240 and 312556.
T.P.R.\ acknowledges support from the European Union's Horizon 2020 research and innovation programme under the Marie Sk{\l}odowska-Curie grant agreement No~838996.
M.K.\ acknowledges funding from Academy of Finland under grant No~295602.
We acknowledge computational resources provided by the CSC -- IT Center for Science (Finland), the Aalto Science-IT project, and the Swedish National Infrastructure for Computing (SNIC) at PDC (Stockholm).
\end{acknowledgments}

\bibliography{references}

\end{document}